\documentclass[aps,prb,onecolumn,preprint]{revtex4-1} 

\usepackage{graphicx}
\usepackage{dcolumn}
\usepackage{bm}
\usepackage{color}
\usepackage{amsmath}
\usepackage{longtable} 

\definecolor{orange}{rgb}{0.9,0.4,0.1}
\definecolor{brown}{rgb}{0.54,0.27,0.07}

\begin{document}

\preprint{Caballero, et al.}

\title{Comparing large lecture mechanics curricula using the Force Concept Inventory:\\ A five thousand student study}

\author{Marcos D. \surname{Caballero}}
	\altaffiliation[Current Address: ]{Department of Physics, University of Colorado at Boulder, Boulder, CO 80309}
	\email{marcos.caballero@colorado.edu}
\author{Edwin F. \surname{Greco}}
\author{Eric R. \surname{Murray}}
\affiliation{School of Physics, Georgia Institute of Technology, Atlanta, GA 30332}

\author{Keith R. \surname{Bujak}}
\author{M. Jackson \surname{Marr}}
\author{Richard \surname{Catrambone}}
\affiliation{School of Psychology, Georgia Institute of Technology, Atlanta, GA 30332}

\author{Matthew A. \surname{Kohlmyer}}
\affiliation{Advanced Instructional Systems, Inc., Raleigh, NC 27696}

\author{Michael F. \surname{Schatz}}
	\email[Corresponding Author: ]{michael.schatz@physics.gatech.edu}
\affiliation{School of Physics, Georgia Institute of Technology, Atlanta, GA 30332}


\date{\today}

\begin{abstract}
The performance of over 5000 students in introductory calculus-based mechanics courses at the Georgia Institute of Technology was assessed using the Force Concept Inventory (FCI). Results from two different curricula were compared: a traditional mechanics curriculum and the Matter \& Interactions (M\&I) curriculum. Post-instruction FCI averages were significantly higher for the traditional curriculum than for the M\&I curriculum; the differences between curricula persist after accounting for factors such as pre-instruction FCI scores, grade point averages, and SAT scores. FCI performance on categories of items organized by concepts was also compared; traditional averages were significantly higher in each concept. We examined differences in student preparation between the curricula and found that the relative fraction of homework and lecture topics devoted to FCI force and motion concepts correlated with the observed 
performance differences. 
Limitations of concept inventories as instruments for evaluating 
curricular 
reforms are discussed.
\end{abstract}

\maketitle

\section{\label{sec:intro}Introduction}

Each year more than 35\% of American college and university students enroll in a physics course.\cite{sadler2001success} Only a small fraction of these students ultimately complete a degree in physics; the vast majority pursue a degree in engineering or another science.\cite{ies2006degrees}  Many are students in an introductory physics course; approximately 175,000 students each year enroll in introductory calculus-based physics.\cite{mulvey2010focus}  However, many of these students fail to acquire an effective understanding of concepts, principles, and methods from these introductory courses. Rates of failure and withdrawal from these courses are often high and substantial research into this subject has shown that students' misconceptions in physics persist after instruction.\cite{hallouna1985csc, redish1999teaching} This paper describes an attempt to evaluate, using a multiple-choice concept inventory, \cite{hestenes1992fci}
a reformed introductory mechanics curriculum\cite{mandi} which aims to mitigate these issues by altering the goals and content (i.e., the {\bf curriculum}) of the typical mechanics course.


To help improve student learning in physics, 
many new methods of content delivery ({\bf pedagogy}) have been 
developed in recent years.  Typically, these methods have been implemented with little change to course curricula.
Well established pedagogical modifications now used widely include tutorials,\cite{tutorialswash} clicker questions,\cite{wieman_clickers} peer instruction,\cite{mazurpeer}  Socratic tutorial homework systems,\cite{morote2009course} multiple representations of concepts and principles,\cite{brewe2008modeling} and reconfigurations of the instructional environment.\cite{beichner} 
There is ample evidence that students who experience these pedagogical reforms perform better on end-of-course concept inventories than students in passive lecture courses. Concept inventories are useful tools to make such comparisons in these cases where all courses (with and without pedagogical reform) share, for the most part, the same core content and goals.

By contrast, 
there is sparse research on how student learning is affected by substantial alterations to the goals and content (curriculum) of introductory physics courses.  One reason for the lack of such work is the relative absence of alternative introductory physics curricula; improvements to the introductory physics curriculum have not progressed as rapidly as improvements in pedagogy. 
Most students learn introductory physics following a canon of topics that has remained largely unchanged for decades regardless of the textbook edition or authors.   
Moreover, choosing how to compare a novel introductory physics curriculum to a traditional curriculum presents a challenge.
Concept inventories can be used for such a comparison.\cite{bema09}
However, there are a number of issues that are peculiar to curricular comparison including which topics to select for comparison and the alignment of the inventory with the goals and content of the curricularly reformed course.

At the Georgia Institute of Technology (Georgia Tech, GT), we have used a concept inventory to evaluate student understanding of force and motion in 
both a traditionally sequenced introductory calculus-based mechanics course \cite{knight04} and an introductory calculus-based mechanics course using the reform curriculum, Matter \& Interactions (M\&I). \cite{mandi}
While both courses employ similar pedagogical best practices, M\&I differs from the traditional curriculum in its focus on the generality of fundamental physical principles, the introduction of microscopic models of matter, and its coherence in linking different domains of physics. \cite{atomsajp, mechajp} 
In particular, M\&I revises the learning progression of the first semester introductory mechanics course by reorganizing and augmenting the traditional sequence of topics.  
For example, early emphasis is placed on the impulse-momentum theorem (referred to as the ``momentum principle'' in the M\&I curriculum), $\Delta \vec{p} = \vec{F} \Delta t$, with iterative application of the momentum principle over short time steps to predict motion by means of both analytic calculation and numerical 
computation. \cite{computajp, kohlmyer_thesis}
Furthermore, M\&I introduces non-constant forces early on to demonstrate the predictive power of this principle.
By contrast, in a traditional curriculum, early emphasis is placed on study of the kinematics of special case situations (e.g., motion under constant acceleration) without explicit discussion of dynamics.
Further discussion of differences between the M\&I curriculum and a traditional curriculum can be found elsewhere. \cite{mandi,atomsajp,mechajp,computajp}


At present, there is no mechanics concept inventory whose force and motion content has been explicitly aligned with goals and content in both courses both traditional and  M\&I mechanics reform. (By contrast, there is at least one concept inventory that is aligned with both traditional and M\&I 
electromagnetism. \cite{dingBEMA,bema09}) Under these circumstances, we chose the Force Concept Inventory (FCI) to make a comparative evaluation both because it is widely used and because of anecdotal evidence of underperformance on the FCI in courses using M\&I mechanics at other institutions.
The FCI was designed to probe performance on force and motion in a particular way; within the context of specific situations, FCI questions were designed to draw out common misconceptions and naive notions about force and motion.\cite{hestenes1992fci} 
As a result, a measurement of performance using the FCI does not provide a comprehensive picture of student understanding of force and motion; the nuances of interpreting student performance on the FCI have been well-documented.\cite{huffman1995dfc, hestenes1995ifc, heller1995ifc, steinberg1997pmc, rebello2004eds} 
Furthermore, the FCI was not specifically developed to compare student performance between courses, but has been used for this purpose.\cite{hake6000}
Thus, to emphasize the idea the FCI probes force and motion in a restricted way, we indicate, in this paper, the content of and concepts covered by the FCI as {\it FCI force and motion concepts}. \footnote{To obtain a copy of the FCI, contact David Koch (ASU) by email: {\tt FCIMBT@verizon.net}}
Moreover, we qualify all of our comparative measures with the understanding that the FCI was designed in the context of a traditional sequenced curriculum before the M\&I curriculum came into existence.

The description of our study is presented below as follows: In Sec. \ref{sec:narrative}, we describe the organizational structure of the Georgia Tech mechanics courses.  Sec. \ref{sec:summary} summarizes the results of the in-class testing. In Sec. \ref{sec:item_analysis}, we present an analysis of FCI performance by individual item and concept. Sec. \ref{sec:origins} examines possible reasons for performance differences observed in Secs. \ref{sec:summary} and \ref{sec:item_analysis}.  In Sec. \ref{sec:discussion}, we provide more insight into the performance differences, make concluding remarks, and outline possible future research directions.

\section{\label{sec:narrative}Introductory Mechanics at Georgia Tech}

\begin{longtable}{|c|c|c|c|c|c|c|}
\caption{Georgia Tech FCI test results are shown for twenty-two traditional sections (T1-T22) and six Matter \& Interactions sections (M1-M6). Different lecturers are distinguished by a unique letter in column L. The average FCI score $O$ for $N_O$ students completing the course is shown for all sections. Moreover, in those sections where data are available, the average FCI score $I$ for $N_I$ students entering the course are indicated.
$N_m$ is the number of students in a given section who took the FCI both at the beginning and at the end (i.e., matched data) of their mechanics course.}\label{tab:summary_data}\\\hline
{\bf ID} & {\bf L} & {\bf I\%} & {\bf N$_{\mathrm{I}}$} & {\bf O\%} &  {\bf N$_{\mathrm{O}}$} & {\bf N$_{\mathrm{m}}$}\\ \hline
\endhead
T1 & A & 49.95$\pm$3.05 & 194 & N/A & N/A & N/A\\\hline
T2 & A & 52.13$\pm$2.80 & 208 & N/A & N/A &  N/A\\\hline
T3 & B & 51.76$\pm$2.88 & 207 & N/A & N/A &  N/A\\\hline
T4 & B & 51.39$\pm$2.91 & 196 &  N/A & N/A &  N/A\\\hline
T5 & C & 46.39$\pm$2.69 & 205 & N/A & N/A &  N/A\\\hline
T6 & D & 45.83$\pm$3.53 & 139 & 70.13$\pm$3.60 & 103 & 97\\\hline
T7 & C & 47.27$\pm$2.86 & 182 & 64.01$\pm$3.05 & 158 & 139\\\hline
T8 & C & 42.03$\pm$2.55 & 194 & 61.26$\pm$3.14 & 140 & 133\\\hline
T9 & A & 52.16$\pm$2.99 & 182 & 73.44$\pm$2.97 & 127 & 122\\\hline
T10 & A & 48.12$\pm$2.72 & 188 & 73.97$\pm$2.92 & 116 & 113\\\hline
T11 & B & 49.82$\pm$2.88 & 182 & 75.35$\pm$3.48 & 104 & 98\\\hline
T12 & B & 49.58$\pm$3.43 & 168 & 72.04$\pm$4.06 & 93 & 88\\\hline
T13 & E & 52.81$\pm$3.25 & 141 & 77.20$\pm$3.38 & 88 & 84\\\hline
T14 & E & 40.36$\pm$2.65 & 183 & 67.33$\pm$3.53 & 140 & 132\\\hline
T15 & F & 46.39$\pm$3.05 & 180 & 69.59$\pm$3.36 & 131 & 120\\\hline
T16 & F & 40.74$\pm$2.84 & 194 & 65.22$\pm$3.60 & 115 & 108\\\hline
T17 & E & 48.02$\pm$3.17 & 160 & 71.82$\pm$3.57 & 121 & 109\\\hline
T18 & A & 50.19$\pm$3.05 & 175 & 74.05$\pm$3.44 & 107 & 105\\\hline
T19 & A & 53.49$\pm$3.37 & 174 & 72.10$\pm$3.52 & 103 & 94\\\hline
T20 & E & 53.36$\pm$3.27 & 143 & 78.52$\pm$3.68 & 97 & 89\\\hline
T21 & B & 49.43$\pm$3.00 & 180 & 75.79$\pm$3.12 & 121 & 115\\\hline
T22 & B & 51.48$\pm$3.09 & 182 & 79.92$\pm$2.81 & 119 & 116\\\hline
M1 & G & N/A & N/A & 35.71$\pm$5.62 & 28 & N/A\\\hline
M2 & H & 54.12$\pm$3.86 & 127 & 64.68$\pm$4.16 & 116 & 111\\\hline
M3 & G & 45.01$\pm$3.11 & 145 & 56.49$\pm$3.38 & 148 & 133\\\hline
M4 & H & 45.57$\pm$3.51 & 143 & 62.27$\pm$3.37 & 141 & 128\\\hline
M5 & I & 45.35$\pm$3.61 & 134 & 62.70$\pm$3.44 & 132 & 110\\\hline
M6 & J & 44.83$\pm$2.50 & 214 & 54.15$\pm$3.06 & 196 & 180\\\hline
\end{longtable}

The typical introductory mechanics course at Georgia Tech is taught with three one-hour lectures per week in large lecture sections (150 to 250 students per section) and three hours per week in small group (20 student) laboratories and/or recitations. 
Attendance of lecture sections is optional but encouraged through a small incentive (2-5\% of course grade). 
Attendance of laboratory and recitation sections is mandatory. 
In the traditional (TRAD) curriculum, each student attends a two-hour laboratory and, in a separate room, a one-hour recitation each week.\footnote{The use of the label ``traditional'' to describe the non-M\&I course is a matter of convenience. Georgia Tech's traditional course is pedagogically reformed, however, the traditional course explores the typical content and examples presented in most introductory physics courses.}
In the M\&I curriculum, students meets once per week with teaching assistants for a single three-hour laboratory/recitation session involving both lab activities (for approximately 2 hours on average) and separate recitation activities (for approximately 1 hour on average). 
Room/TA scheduling is responsible for the differences in instructional locations between the two courses.
The student population of the mechanics course (both traditional and M\&I) consists of approximately 
85\% engineering majors and 
15\% science (including computer science) majors.

Table \ref{tab:summary_data} summarizes the FCI test results for individual sections. In most traditional (T6-T22) and all M\&I sections, $N_O$ students in each section took the FCI during the last week of class at the completion of the course. In all of the traditional sections and in the majority of M\&I sections (M2-M6), $N_I$ students in each section took the FCI at the beginning of the course during the first week of class. For a given section, $N_I$ is approximately equal to the number of students enrolled in that section. $N_O$ is usually smaller than $N_I$, sometimes substantially so (e.g., T12, T13 and T20). M\&I students took both the pre- and post-test during their required laboratory section. Students of the traditional curriculum typically took the pre-test during the first lecture or lab section. Traditional students were asked to attend an optional section during their evening testing period to take the post-test. Students become busy with other coursework near the end of the semester, hence fewer traditional students attended this optional evening section. 
In each section, only those $N_m$ students who took the FCI both on entering and on completion of the course are considered for the purposes of computing any type of gain (Sec. \ref{sec:summary}). The FCI was administered using the same time limit (30 minutes) for both traditional and M\&I students. M\&I students were given no incentives for taking the FCI; they were asked to take the exam seriously and told that the score on the FCI would not affect their grade in the course. Traditional students taking the FCI were given bonus credit worth up to a maximum of 0.5\% of their final course score, depending in part on their performance on the FCI. This incentive difference between the two curricula has no bearing on the performance differences we observe in our data (Sec. \ref{sec:origins}).

\section{\label{sec:summary}Summary of Measurements from In-class Testing}

\begin{figure}[t]
	\centering
		\includegraphics[width=0.96\linewidth]{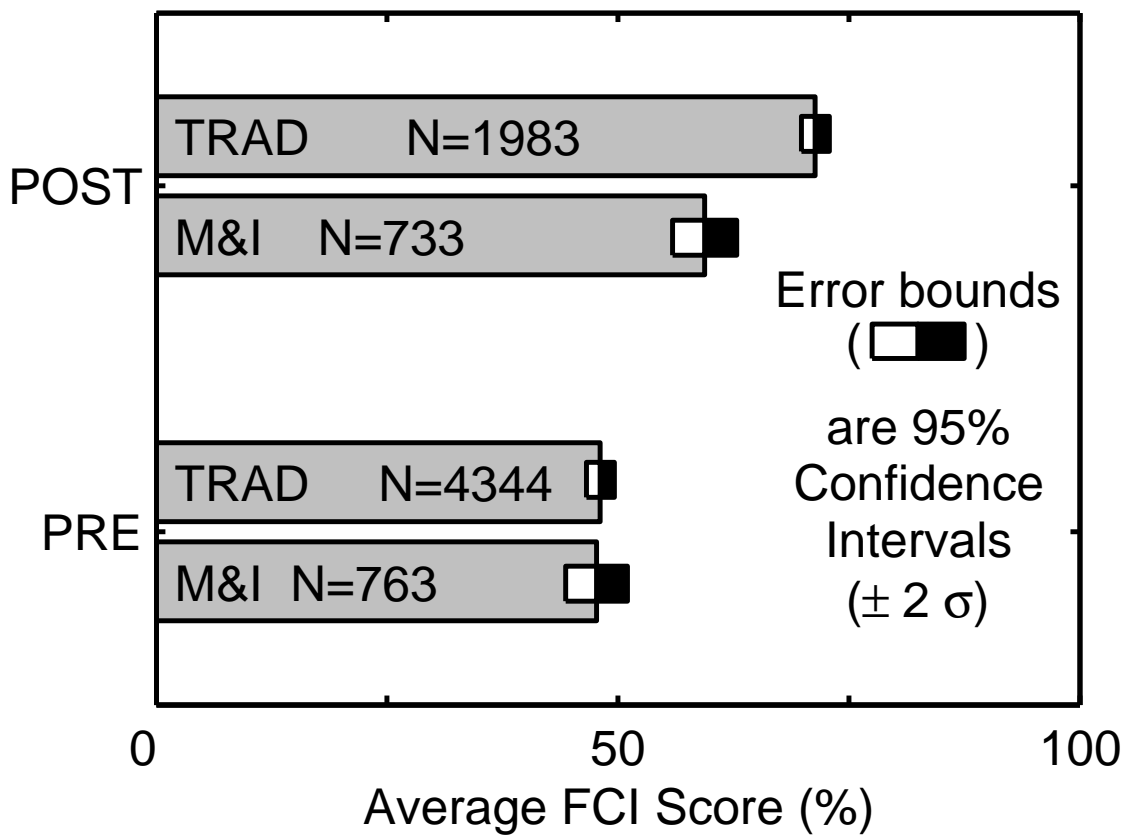}
	\caption{Average pre- and post-instruction FCI scores at Georgia Tech. The average FCI pre- and post-test scores are shown for students who took a one-semester mechanics course with either the traditional (TRAD) or Matter \& Interactions (M\&I) curriculum. 
The number of students ($N$) tested for each curriculum is indicated in the figure. The error bounds represent the 95\% confidence intervals (estimated from the t-statistic) on the estimate of the average score.}\label{fig:gt_summary}
\end{figure}

\begin{figure}[t]
	\centering
		\includegraphics[width=0.96\linewidth]{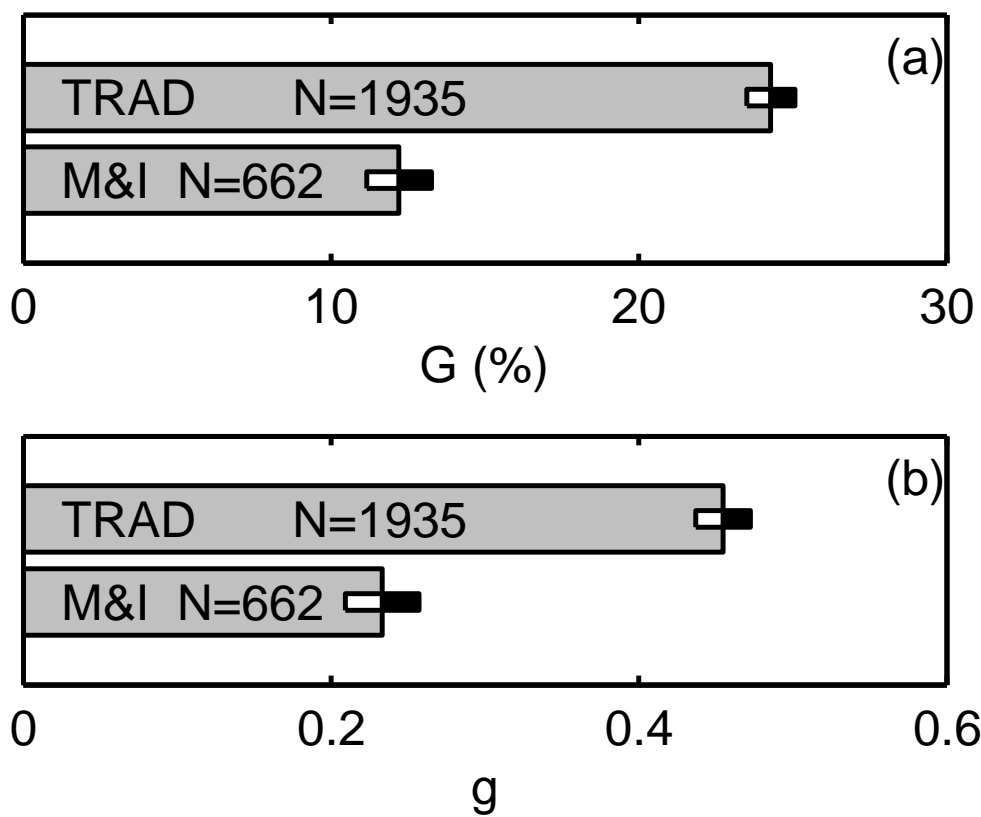}
	\caption{Gain in understanding of mechanics as measured by the FCI. The increase in student understanding
resulting from a one-semester traditional (TRAD) or Matter \& Interactions (M\&I) course is measured using (a) the average raw gain $G$ and (b) the average normalized gain $g$. 
Only students with matched scores were used for this figure (see Table \ref{tab:summary_data}). The error bounds represent the 95\% confidence intervals (estimated from the t-statistic) on the estimate of (a) the raw gain and (b) the normalized gain.}\label{fig:gains}
\end{figure}

\begin{figure}[t]
	\centering
		\includegraphics[width=0.96\linewidth]{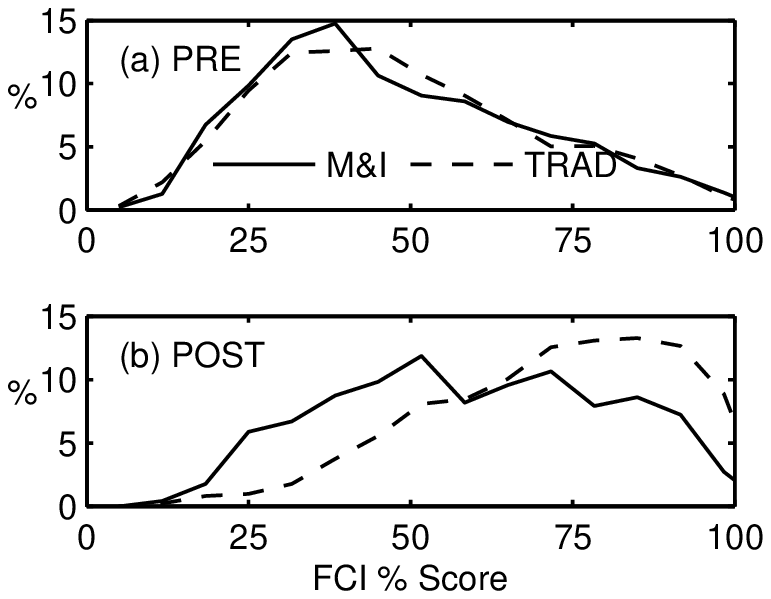}
	\caption{FCI score distributions by curriculum. The distribution of FCI test scores for students before (a) and after (b) completing a mechanics course with either a traditional (dashed line) or M\&I curriculum (solid line) are shown. The total number of students tested in each curriculum is the same as in Fig. \ref{fig:gt_summary}. The plots are constructed from binned data with bin widths equal to approximately 6.7\% of the maximum possible FCI score (100\%). }
\label{fig:gt_prepost}
\end{figure}

The FCI pre-test scores for Matter \& Interactions (M\&I) and traditional students did not differ significantly (mean FCI score, 48.9\% for TRAD vs. 47.4\% for M\&I). 
By contrast, on the FCI  post-test, traditional students significantly outperformed M\&I students (mean FCI score, 71.3\% for TRAD vs. 59.3\% for M\&I). In Fig. \ref{fig:gt_summary}, these mean scores have been reported with 95\% confidence intervals estimated from the t-statistic for each distribution. \cite{zhou_biostats}
A common measure of the change in performance from pre-test to post-test \cite{hake6000} is the average percentage gain, $G = (O - I)*100\%$, where $I$ is the average fractional FCI score for students entering a mechanics course, and $O$ is the average end-of-course fractional FCI score. We also report an average normalized gain $g$, where $g = (O-I)/(1 - I)$, and where $(1 - I)$ represents the maximum possible fractional gain that could be obtained by a class of students with an average incoming fractional FCI score of $I$. For the gains reported in Fig. \ref{fig:gains}, 95\% confidence intervals have been estimated from the t-statistic for the distributions of $G$ and $g$. The data are shown for $N_m$ students (Table \ref{tab:summary_data}).



FCI pre-test score distributions were found to be statistically indistinguishable between the two curricula, which is evident from Fig. \ref{fig:gt_prepost}(a). By contrast, distributions of post-test FCI scores were dissimilar; the traditional distribution was shifted towards higher scores (Fig. \ref{fig:gt_prepost}(b)). This is consistent with the finding that the mean score achieved by traditional students were higher than their M\&I peers on the post-test (Fig. \ref{fig:gt_summary}). 
Because the distributions of FCI pre- and post-test scores were non-normal, the similarity of the distributions was compared using a rank-sum test. \cite{conover_nonpara,nonparabook}

An examination of measures of student performance entering each course suggests that the incoming and outgoing student populations of both curricula were identical. 
We obtained and examined students' grade point averages (GPA) upon entering the mechanics course, SAT Reasoning Test (SAT) scores, and the grades earned in the mechanics course; we found no significant difference in the distributions of any of these metrics using a rank-sum test.



Mean scores differed between one or more sections within a given curriculum as measured by a Kruskal-Wallis test. \cite{nonparabook} Given this section effect, we compared the three lowest performing traditional sections (T7, T8, \& T16) to the three highest performing M\&I sections (M2, M4, \& M5) to determine if this section effect enhanced the overall observed differences in the normalized gains. Post-test FCI scores were statistically indistinguishable between these subsets. 
However, traditional students in these sections had significantly lower pre-test FCI scores. 
Hence, students in these lower performing traditional sections achieved significantly higher normalized gains. 
We also compared the FCI post-test scores achieved by the three traditional sections with lowest normalized gains (T14, T18, \& T22) to the M\&I sections with the highest normalized gains (M3, M4, \& M5). 
Pre-test FCI scores were significantly higher for the M\&I subset 
while post-test scores were higher for the traditional subset. 
Thus normalized gains achieved by traditional students in this subset were higher.
\section{\label{sec:item_analysis}Item Analysis of FCI Measurements}

\begin{figure}
	\centering
		\includegraphics[width=1\linewidth, clip, trim = 0mm 15mm 26mm 15mm]{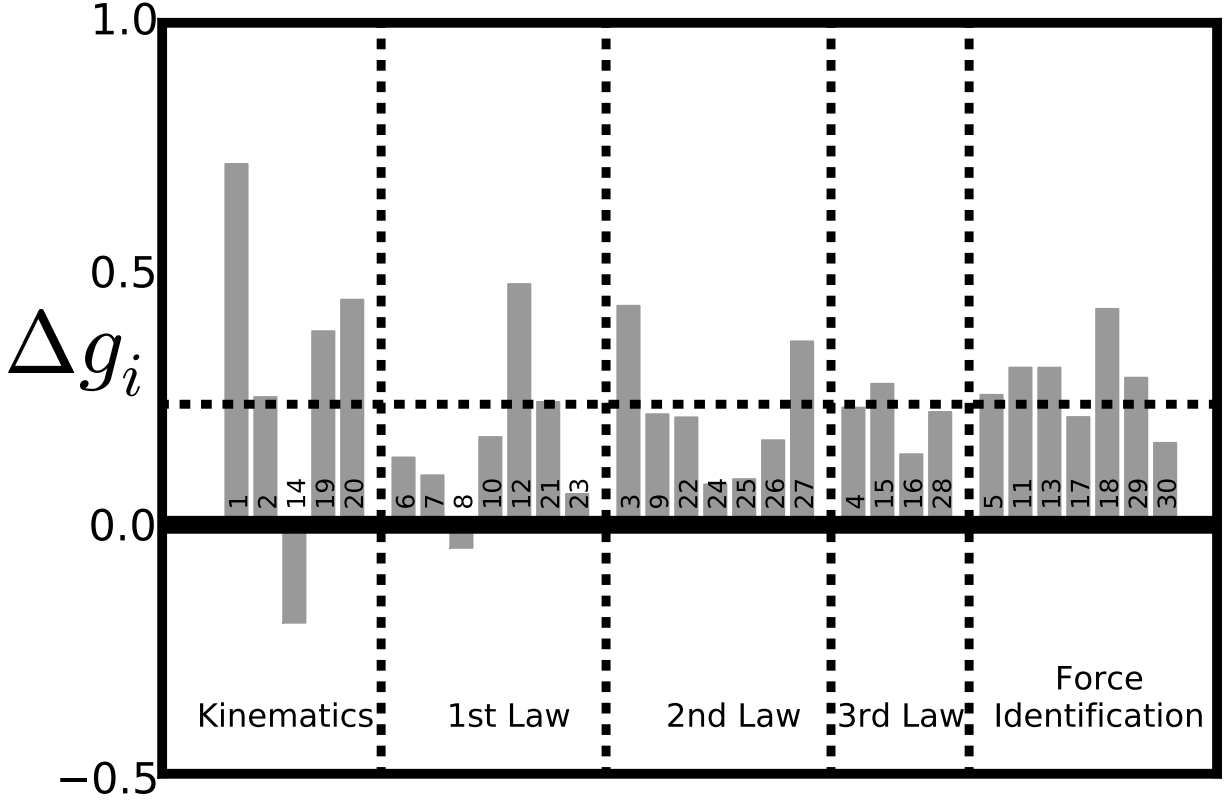}
	\caption{Difference in performance for individual FCI items and mechanics concepts. The difference in performance ${\Delta g_{i}}$ between traditional and M\&I students is shown for each question on the FCI. Positive (negative) ${\Delta g_{i}}$ indicates superior performance by traditional (M\&I) students on individual questions. The numerical labels indicate the corresponding question number in order of appearance on the FCI. The items are grouped together into one of five concepts: Kinematics,  Newton's first law, Newton's second law, Newton's third law, and Force Identification. The horizontal line (dash) illustrates the value of $\overline{\Delta{g}}$ the mean difference in the item gains between curricula.}\label{fig:pcanew}
\end{figure}

Student performance on individual questions or groups of questions was used to determine on which FCI force and motion concepts students in the traditional curriculum outperformed M\&I students. 
Questions on the FCI were sorted into concept categories using Hestenes' original conceptual dimensions, \cite{hestenes1992fci} but we required that each question be placed in only one category. 
In our work only five concept categories were used: Kinematics, Newton's 1st Law, Newton's 2nd Law, Newton's 3rd Law, and Force Identification. 
The first four of these categories were identical to Hestenes' dimensions and Force Identification was a renamed category which contained questions from Hestenes' Kinds of Forces dimension. 
In Fig. \ref{fig:pcanew} the items that comprise each category are listed. 
Note that this was an \emph{a priori} categorization based on our judgment of the concepts covered by the items; it is not the result of internal correlations or factor analysis based on student data.


We used the normalized gain in performance on a per question basis to quantify item performance. 
We define an item gain, 
\begin{equation}{\label{eqn:itemg}}
{g}_{i} =  \frac{f_{\mathrm{post},i}-f_{\mathrm{pre},i}}{1-f_{\mathrm{pre},i}}
\end{equation}
where $f_{pre,i}$ and $f_{post,i}$ are the fraction of students responding correctly to the $i^{th}$ item on the pre- and post-test respectively. 
This measure normalizes the gain in performance on a single item by the largest possible gain given the students' pre-test performance on that item; ${g}_{i}$ is essentially the Hake gain for a single item. 
To discern which questions have large item gains, we can compare $g_{i}$ for each question to the mean item gain, 
\begin{equation}{\label{eqn:avgg}}
\bar{g} = \frac{1}{N} \sum_i^N g_i
\end{equation}
where $N$ is the number of items on the FCI.

To illustrate the differences between curricula succinctly, we computed difference in normalized item gains between the two curricula. We define the difference in normalized item gain, 
\begin{equation}
\Delta g_{i} = {g}_{i}^{T} - {g}_{i}^{M}
\end{equation}
where ${g}_{i}^{T}$ and ${g}_{i}^{M}$ are the normalized gain for the $i^{th}$ item achieved by traditional and M\&I students respectively. 
We discovered on which questions students' item gains in each curricula differed the most by comparing $\Delta g_{i}$ for each item to the mean difference in the item gains between curricula, 
\begin{equation}
\overline{\Delta{g}} = \frac{1}{N}\sum_i^N \Delta g_i.
\end{equation}

The plot of ${\Delta g_{i}}$ illustrates better performance by traditional students across all concepts on the FCI (Fig. \ref{fig:pcanew}). We observed that $\Delta{g_{i}}$ is positive for almost all questions, and 45\% of the questions had values of $\Delta{g_{i}}$ greater than $\overline{\Delta g} = 0.238$. The grouping of the FCI questions by category permits one to visualize which concepts contributed most 
strongly to the difference in performance. For example, the difference in performance on the Force Identification concept was striking, where 5 of the 7 questions in this category had $\Delta g_i > \overline{\Delta g}$.

Moreover, this grouping helps one to determine on which concepts differences in item gains were greatest. We computed the difference in the average concept gain, 
\begin{equation}\label{eqn:avgcg}
\overline{\Delta {g}_c} = \frac{1}{N_c} \sum_{i \epsilon c} \Delta g_{i}
\end{equation}
where $N_c$ is the number of items covering concept $c$. Concepts with higher $\overline{\Delta {g}_c}$ were those on which traditional students achieved higher normalized gains than M\&I students. The Kinematics and Force Identification concepts had the highest values of $\overline{\Delta {g}_c}$ (shown in Table \ref{tab:deltag}). By contrast, we found $\overline{\Delta {g}_c}$ for Newton's 1st Law which was well below $\overline{\Delta {g}}$. The remaining two concepts had values of $\overline{\Delta {g}_c}$ slightly below $\overline{\Delta {g}}$.

\begin{table}
\caption{The average difference in item gains between curricula are computed for the items in each FCI force and motion concept, $\overline{\Delta {g}_c}$. Each $\overline{\Delta {g}_c}$ is positive, indicating better average item gains for traditional students across all FCI force and motion concepts. Concepts with higher $\overline{\Delta {g}_c}$ are those for which traditional students achieve higher normalized gains than M\&I students. Traditional students achieve the highest values of $\overline{\Delta {g}_c}$ on the Kinematics and Force Identification concepts and lowest on Newton's 1st Law concept. The measures are presented along with their variance.}
\begin{tabular}{|l|c|c|}\hline
{\bf FCI force and motion concepts \hspace*{40pt}} & {${\overline{\Delta {g}_c}}$} & {$\sigma^2$}\\\hline
Kinematics & 0.32 & $<$0.01\\ 
Newton's 1st Law & 0.16 & $<$0.01\\
Newton's 2nd Law & 0.22 & $<$0.01\\
Newton's 3rd Law & 0.22 & 0.01\\
Force Identification & 0.28 & 0.05\\\hline
\end{tabular}
\label{tab:deltag}
\end{table}

\section{\label{sec:origins}Contributions to the Performance Differences}



We turn now to the examination of factors that might contribute to higher FCI post-test scores by traditional students, including grade incentives, differences in pedagogy, and differences in instruction (e.g., homework and lecture topics).  




The incentive given to traditional students to take the FCI was too small to account for the marked differences in performance indicated in Figs. \ref{fig:gt_summary}, \ref{fig:gains}, and \ref{fig:gt_prepost}(b).  As mentioned earlier (Sec. \ref{sec:narrative}), traditional students were provided with an incentive to take the FCI while M\&I students received no incentive. In principle, sufficiently large incentives can impact FCI outcomes. For example, Ding, et al. found a 10-15\% increase in FCI post-test scores if scores on the FCI were valued as highly as course exams. \cite{ding4etc}  To check for this incentive effect, we offered similar incentives (i.e., a maximum of 0.5\% bonus to overall course grade) to both traditional and M\&I students who took the FCI post-test at Georgia Tech in the fall of 2009. During this term, we found the performance differences for M\&I and traditional students were similar to those reported in this paper.
FCI data from fall 2009 was not included in this paper because instructional changes had been made to the M\&I course; M\&I sections M1-M5 had similar homework exercises, lectures, and laboratories.



The performance differences cannot be attributed to differences in pedagogy. It is well-known that using interactive engagement (i.e., ``clicker'' questions, ConceptTests, Peer Instruction, etc.) can improve students' conceptual understanding in introductory and advanced courses. \cite{hake6000,crouch2001peer,pollock-use} However, all sections (both traditional and M\&I) were largely indistinguishable with respect to interactive engagement: all sections used similar methods (``clicker'' questions) with similar intensity (3-6 ``clicker'' questions per lecture period). 






We examined whether differences in coursework (homework) could be connected to performance differences on the FCI. We categorized the 575 traditional homework questions and the 756 M\&I homework questions. Questions were placed into one or more categories depending on the topical nature of the problem and the principles needed to answer the question. Categories included the five FCI force and motion concepts discussed in Sec. \ref{sec:item_analysis} as well as several other concepts which do not appear on the FCI (e.g., Angular Momentum). The {\it Kinematics category} included questions about the relationships between position, velocity, and acceleration that did not refer to the underlying dynamical interactions that cause changes in these quantities. Questions in the {\it Newton's 1st Law category} included qualitative questions which discussed the direction of motion and its relationship to applied forces. The {\it Newton's 2nd Law category} included questions with a heavy emphasis on contact forces and resolving unknown forces, but excluded open-ended questions in which the prediction of future motion is the goal (e.g., using iterative methods to predict the motion of an object). Questions in the {\it Newton's 3rd Law category} included questions in which Newton's 3rd law was treated as an isolated law, that is, where there was no reference to the underlying reciprocity of long range electric interactions which causes it. Generally, it was applied to contact forces and gravitational interactions. The {\it Force Identification category} included questions in which the direction and relative strength of forces acting on a body or set of bodies were represented by diagrams (i.e., force-body diagram).  
The aforementioned categories represent those concepts that are covered extensively in the first half of a traditional physics course and were heavily represented on the FCI. 

\begin{table}
\caption{An estimate of the fraction of homework questions covering a particular FCI concept in the two mechanics curricula is compared. Subtopics for these homework questions were not mutually exclusive. The relative fraction of homework questions covering FCI force and motion concepts and some individual FCI concepts (i.e., Kinematics, Newton's 2nd Law, Newton's 3rd Law, and Force Identification) is greater in the traditional curriculum.}
\begin{tabular}{|l|c|c|}\hline
{\bf Est. Fraction of HW Questions \hfill} & {\bf M\&I} & {\bf TRAD}\\\hline
FCI force and motion concepts & 0.26 & 0.57 \\\hline
\multicolumn{3}{|l|}{\bf HW Subtopics (not exclusive)} \\\hline
Kinematics & 0.10 & 0.26\\ 
Newton's 1st Law & $<$0.01 & $<$0.01\\
Newton's 2nd Law & 0.15 & 0.25\\
Newton's 3rd Law & $<$0.01 & 0.04\\
Force Identification & 0.01 & 0.11\\\hline
\end{tabular}
\label{tab:hwcat}
\end{table}

The difference in the relative fraction of homework questions covering FCI force and motion concepts between the curricula (Table \ref{tab:hwcat}) reflect the overall performance differences observed in Figs. \ref{fig:gt_summary} - \ref{fig:gt_prepost}. Furthermore, the differences in the relative fractions of homework questions corresponding to individual FCI concepts were consistent with the results from our item analysis (Fig. \ref{fig:pcanew}). The relative fraction of homework questions was computed by first categorizing questions, then counting the number of questions covering the concepts of interest and dividing by the total number of homework questions given in a curriculum. The relative fraction of homework questions covering FCI force and motion concepts differed by more than a factor of 2 in favor of the traditional curriculum. 
On individual FCI concepts, we found a lower relative fraction of homework questions in the M\&I curriculum compared to the traditional curriculum on four of the five concepts: Kinematics, 
Newton's 2nd Law, 
Newton's 3rd Law, 
and Force Identification. 
On most FCI questions about these concepts traditional students outperformed M\&I students (Sec. \ref{sec:item_analysis} \& Fig. \ref{fig:pcanew}). 
We found that the relative fraction of Newton's 1st Law questions were similar.
This signature was also observed in our item analysis (Fig. \ref{fig:pcanew}); the Newton's 1st Law FCI concept had the smallest $\overline{\Delta {g}_c}$ (Sec.\ref{sec:item_analysis}).

The difference in the relative fraction of force and motion lectures/readings between the curricula (Table \ref{tab:readcat}) was consistent with the overall performance differences observed in Figs. \ref{fig:gt_summary}, \ref{fig:gains}, \& \ref{fig:gt_prepost}(b). 
The relative fraction of lectures/readings which cover FCI force and motion concepts was greater by nearly a factor of 2 for the traditional curriculum. 
This result is consistent with the difference in the relative fraction of homework questions (Table \ref{tab:hwcat}).
However, the differences in the relative fractions of lectures/readings corresponding to individual FCI concepts showed mixed results when compared to our item analysis (Fig. \ref{fig:pcanew}).
The relative fractions for three of five concepts were greater for the traditional curriculum: Kinematics, 
Newton's 3rd Law,
and Force Identification.
But on two concepts, the relative fractions of lectures/readings were roughly similar: Newton's 1st Law 
and Newton's 2nd Law. 
Lecture and reading topics were examined and categorized for each curriculum using the same categories as our homework question analysis.

\begin{table}
\caption{An estimate of the fraction of lecture/reading topics in the two mechanics curricula is compared. Subtopics for these lectures/readings were not mutually exclusive. The relative fraction of lectures/readings in the traditional course is greater for the Kinematics, Newton's 3rd Law, and Force Identification topics which is consistent with their superior performance in those concepts on the FCI. However, on Newton's 1st and 2nd Laws, the relative fraction of lectures/readings are roughly similar.}
\begin{tabular}{|l|c|c|}\hline
{\bf Est. Fraction of Lecture Topics \hfill} & {\bf M\&I} & {\bf TRAD}\\\hline
FCI force and motion concepts & 0.26 & 0.44 \\\hline
\multicolumn{3}{|l|}{\bf Lecture Subtopics (not exclusive)} \\\hline
Kinematics & 0.07 & 0.21\\ 
Newton's 1st Law & 0.02 & 0.01\\
Newton's 2nd Law & 0.09 & 0.08\\
Newton's 3rd Law & 0.01 & 0.03\\
Force Identification & 0.06 & 0.11\\\hline
\end{tabular}
\label{tab:readcat}
\end{table}

\section{\label{sec:discussion}Closing Remarks and Lessons Learned}

We have found that students who completed an introductory mechanics course which employs the Matter \& Interactions curriculum earned lower post-test FCI scores than students who took a traditionally sequenced curriculum. 
The differences in performance were significant and were supported by the number of students involved in the measurement. 
We demonstrated that these differences cannot be explained by differences in the incoming or outgoing population of students between the courses (i.e., SAT scores, GPA, etc.). 
The overall performance differences between the curricula on the post-test were consistent with the amount of instruction within each curriculum. 
The relative fraction of FCI force and motion concepts that appeared on students' homework and in their lectures was roughly twice as large for the traditional curriculum (Tables \ref{tab:hwcat} \& \ref{tab:readcat}). 
We observed this signature in the differences of the means and distributions of FCI scores (Fig. \ref{fig:gt_summary}, \ref{fig:gains}, \& \ref{fig:gt_prepost}(b)) as well as the average item gain (Eqn. \ref{eqn:avgg}). 
The average item gain for traditional students was roughly twice as large when compared to M\&I students (Sec. \ref{sec:item_analysis}).
Furthermore, we found that traditional students outperformed M\&I students across all subtopics on the FCI (Fig. \ref{fig:pcanew}) and that these differences were consistent with the amount of instruction on individual FCI force and motion concepts that appeared on students' homework (Table \ref{tab:hwcat}). 

{These results indicate the challanges that arise when concept inventories are used to make comparative evaluations of curricular course reforms; such challenges, it should be emphasized, do not typically arise when concept inventories are used to evaluate pedagogical reforms, which often do not affect core course content.   There are at least two considerations that must be kept explictly in mind for the case of curricular reform. First, sensible comparison between courses with and without reform can be made only on content that is present both in courses.  Comparing student performance on curricular materials 
exclusive to one or the other course (e.g., computation in the case of M\&I mechanics) makes 
little sense. Substantial content on force and motion is found in both traditional and M\&I curricula; however, as was mentioned in Sec. \ref{sec:intro}, the specifics of force and motion content differ substantially between the two curricula.  Second, the composition of the evaluation instrument itself 
represents a particular selection of content and goals.  Ideally, for a comparative 
evaluation, the content in the instrument should be aligned with content present in both 
courses (with and without reform) under study; moreover, the goals evaluated by the instrument should be clearly connected to the learning goals of both courses.  Our results support the idea that the content of the FCI is more closely aligned with the content of traditional courses than with M\&I mechanics content, thereby posing significant barriers to interpreting the meaning of FCI performance differences between traditional and M\&I courses.  We emphasize that these difficulties were not present in earlier work by some of us that used a concept inventory (the Brief Electricity and Magnetism Assessment (BEMA)) 
to evaluate comparatively traditional and M\&I curricula for introductory electromagnetism. \cite{bema09}  In the earlier work, comparisons were made based on similar electromagnetism content 
present (in approximately equal measure) in 
both courses; moreover, the instrument used was carefully constructed to align with the minimal subset of content and goals in all courses. \cite{dingBEMA}
}

Notwithstanding its use to evaluate comparatively different curricula, data from the FCI might be used to adjust the 
content and goals of a given curriculum.  For example, if faced with FCI performance similar to that reported here, M\&I mechanics 
course instructors may make the (reasonable) decision that students should have more practice with qualitative questions on topics covered by FCI.  In recent 
terms, we have made small modifications to the M\&I curriculum by adding some homework problems and lab activities that are more aligned 
with the scope of the FCI.  As a consequence we have observed small improvements to the FCI scores of M\&I students.  We have not made
a systematic study of which modifications to the M\&I curriculum are most effective for improving student performance on the FCI.

The main purpose of an evaluation tool is to help answer the questions:  Is the reform doing any good and, if so, is the good worth it?   When the reform is curricular, concept inventories may be used to answer these questions when content and goals are shared by both curricula (with and without reform) with approximately equal intensity. \cite{bema09}  In the absence of this alignment, concept inventories might be used to give some insight into whether anything is ``lost'' with respect to overlaping content and goals.   However, concept inventories (used to make comparisons) 
fail to measure what are perhaps the most interesting aspects of reform: the non-overlapping content and goals that are simply not present in courses without reform. In the case of M\&I mechanics, examples of non-overlapping material includes both new goals (e.g., relating macroscopic physics to microscopic models)\cite{atomsajp} and new content (e.g., computation).\cite{computajp}  There is a need to develop tools to help weigh the gains of a particular reform, so that instructors faced with multiple curricular choices can make informed decisions about which concepts, principles, and methods should be 
included or excluded, emphasized or de-emphasized during the finite time available to them to teach the course.

\begin{acknowledgments}
We would like to thank Andrew Scherbakov and Robert Hume (Office of Minority Education and Development) for their efforts in collecting and organizing the demographic data. This work was supported by National Science Foundation's Division of Undergraduate Education (DUE0618519 and DUE0942076).
\end{acknowledgments}

\bibliography{fci}
\bibliographystyle{apsper}

\end{document}